# Deep Learning-Based Approach for Automatic 2D and 3D MRI Segmentation of Gliomas

**Kiranmayee Janardhan, Christy Bobby T**

*Department of Electronics and Communication Engineering, Ramaiah University of Applied Sciences, Bengaluru, India*
*Email: Kiranmayee.j@msruas.ac.in*

Brain tumor diagnosis is a challenging task for clinicians in the modern world. Among the major reasons for cancer-related death is the brain tumor. Gliomas, a category of central nervous system (CNS) tumors, encompass diverse sub-regions. For accurate diagnosis of brain tumors, precise segmentation of brain images and quantitative analysis are required. A fully automatic approach to glioma segmentation is required because the manual segmentation process is laborious, prone to mistakes, as well as time-consuming. Modern techniques for segmenting gliomas are based on fully convolutional neural networks (FCNs), which can use either two-dimensional (2D) or three-dimensional (3D) convolutions. Nevertheless, 3D convolutions suffer from computational costs and memory demand, while 2D convolutions cannot fully utilize the spatial insights of volumetric clinical imaging data. To obtain an optimal solution, it is vital to balance the computational efficiency of 2D convolutions along with the spatial accuracy of 3D convolutions. This balance potentially be realized by developing an advanced model to overcome these challenges. The 2D and 3D models implemented here are based on UNET architecture, Inception, and ResNet models. The research work has been implemented on the BraTS 2018, 2019, and 2020 datasets. The best performer of all the models' evaluation metrics for proposed methodologies offers superior potential in terms of the effective segmentation of gliomas. The ResNet model has resulted in 98.91% accuracy for 3D segmentations and 99.77% for 2D segmentations. The dice scores for 2D and 3D segmentations are 0.8312 and 0.9888, respectively. This model can be applied to various other medical applications with fine-tuning, thereby aiding clinicians in brain tumor analysis and improving the diagnosis process effectively.

## 1. Introduction

Low-Grade Glioma (LGG) and High-Grade Glioma (HGG) represent two distinct forms of brain tumors derived from glial cells, it is crucial in supporting and protecting neurons within the brain. HGGs, including the aggressive Glioblastoma (GBM), proliferate rapidly, challenging total surgical excision due to their invasive nature. Conversely, LGGs, characterized by slower growth and fewer abnormal cells, generally present a better prognosis, although they can affect individuals across various age groups, particularly children and young adults [1]. In efforts to enhance the analysis of brain glioma medical images, the research community has increasingly focused on developing deep learning (DL) methods, acknowledging their superior performance over conventional techniques [2]. Advanced DL algorithms like VGG16, ResNet, and Inception address the complexities of image analysis, offering promising results [3]. The ResNet architecture, for instance, incorporates skip connections to mitigate the vanishing gradient problem, while Inception adapts various kernel sizes within inception modules to capture multiscale features effectively [4], [5].

Within the domain of medicinal MR image segmentation, the UNET model emerges as a prominent DL architecture for brain glioma segmentation, utilizing encoder-decoder networks to map input images to segmentation maps [6]. Recent enhancements to UNET, including attention mechanisms, residual connections, and multimodal input data integration, further enhance its segmentation accuracy [7], [8], [9]. Studies like the multimodal UNET model suggested by [10] underscore the efficacy of UNET models in glioma segmentation, particularly when incorporating attention mechanisms and multimodal input data. Similarly, architectures like Inception-v3 and Inception-v4 have demonstrated success in glioma segmentation tasks, leveraging their convolutional layers with various filter sizes to extract features at different scales [11]. The work by [12] showcases an active DL-based method utilizing Inception-v3 for glioma segmentation and classification, yielding high dice scores and favorable accuracy across multiple datasets. Furthermore, innovative approaches like the Hyperdense Inception 3D UNET (HI-Net) suggested by [13] offer novel solutions to overcome segmentation challenges, such as factorized convolutional layers in residual inception blocks for contextual feature extraction.

Addressing the current limitations in glioma segmentation methods, including accuracy, computational requirements, and performance variability, remains a crucial motivation for ongoing research. Enhancements in segmentation techniques not only contribute to improved treatment planning and patient management but also pave the way for more effective clinical outcomes in the prognosis of brain gliomas [14].

The contributions of this paper include 3D CNN-based models that address the challenge of losing 3D context in a 2D model. CNNs can directly process 3D volumes without the need for conversion to 2D slices, preserving the spatial information in the z-direction. To access the segmentation results of 3D CNNs, a grouping of dice loss as well as focal loss has been calculated. The dice loss emphasizes the boundary regions where the forecasted as well as ground truth masks overlap, while the focal loss addresses class imbalance in the training data. Combining these approaches can overcome the challenges faced in converting 3D models to 2D models in brain glioma segmentation.

## 2. Literature Review

Brain glioma segmentation is a critical task in medical image analysis aimed at delineating tumor boundaries for accurate diagnosis and treatment planning. Traditionally, segmentation methods have predominantly relied on 3D models to process volumetric medical images like MRI scans. However, in certain scenarios, such as when computational resources are limited or when dealing with 2D MRI images, a 2D model may be preferred or required [15]. The transition from 3D to 2D segmentation involves extracting 2D slices from the 3D volume, which can be achieved through various methods, including selecting specific slices or using specialized software tools like 3D Slicer or ITK-SNAP. Once these 2D slices are extracted, they undergo pre-processing before being fed into a 2D segmentation model like a UNET for tumor segmentation [16]. Despite the advantages of 3D models in preserving spatial information, 2D models offer several benefits in terms of computational efficiency and ease of implementation. While using 2D models may result in some loss of information, especially when tumors extend across multiple slices in the 3D volume, proper slice selection, and post-processing techniques can mitigate these challenges and yield accurate segmentation results [17]. For instance, research by Pereira et al. proposes a 3D UNET architecture for brain tumor segmentation, where 2D slices extracted from the 3D volume are used as input to a modified UNET architecture, demonstrating accurate segmentation results on histology images [18].

In recent years, innovative approaches have emerged to automate brain glioma segmentation, leveraging advancements in deep learning and multimodal imaging data. Hamghalam, Lei, and Wang introduced a 3D convolution to 2D convolution MR patch conversion framework, which effectively segments glioma areas in multimodal MRI data by predicting class labels using both global intra-slice datasets as well as local inter-slice datasets [14]. Similarly, their work on a 2D CNN classifier for glioma segmentation demonstrates effective segmentation results by assigning weights for segmentation based on trainable parameters [19]. Additionally, the 2D Double UNET proposed by Webber et al. presents a method for segmenting glioblastoma from multimodal MRI images, achieving high dice similarity coefficients for various tumor regions [20].

Dong et al. suggested a 2D CNN technique for brain glioma segmentation, converting 3D MR scans into a series of 2D slices and utilizing multiple MRI contrasts for training the network, resulting in promising segmentation accuracy and computational efficiency [21]. Their work introduces the 3D-CNN-Inception architecture, incorporating inception modules to capture multiscale features from input slices, demonstrating the key outcomes on a dataset of glioma patients. Similarly, Russo, Liu, and Di Ieva proposed a method for glioma segmentation in multimodal MRI based on 2D analysis, combining machine learning with human expert knowledge to accurately segment gliomas from MRI images [22]. Their approach involves converting 3D MRI images to 2D images, extracting features using a CNN, and segmenting gliomas using a random forest classifier. Despite the advancements in 2D glioma segmentation, challenges persist in converting 3D models to 2D, including potential information loss, variability in slice thickness, and image quality inconsistencies [18]. These challenges underscore the importance of continued research and optimization efforts in brain glioma segmentation to improve accuracy and efficiency in clinical practice. Overall, the integration of deep learning techniques and multimodal imaging data holds promise for advancing automated brain glioma segmentation, facilitating more accurate diagnosis and treatment planning for patients [15].

The state-of-the-art studies discussed in this section have various limitations such as high computing costs, and huge memory requirements, which might limit their scalability. Furthermore, the existing methods can struggle with accurately segmenting tumors in images with lower contrast and noise problems. Additionally, the explored models are prone to overfitting, where these models can perform well on training data, nevertheless poorly on unseen data. In summary, our work distinguishes through the 3D CNN-based models which address the threats of losing the 3D context within a 2D model. CNNs models manage 3D volumes directly, thereby, maintaining spatial information along the z-axis without converting to 2D slices. Our approach integrating UNET architecture, Inception, and ResNet models. This model successfully attains an improved balance between computational efficiency as well as spatial accuracy, thereby, making it exceptionally well-suited for addressing the complexities of volumetric clinical imaging data and challenges in the segmentation of gliomas.

## 3. Methodology

A. Datasets and Pre-processing

In this work, multi-sequential MRI scans with gliomas of BraTS' 2018, 2019, and 2020 datasets compiled from 19 different MRI imaging facilities [23] are used. The databases, include FLAIR, T1, T1CE, and T2 multisequence MRI scans of GBM and LGG as shown in Table 1.

Each type of scan highlights the glioma in different perceptibility. The raw images are in .nii format and cannot be used directly. Thus, in the pre-processing of data, the .nii images format is converted to a normalized NumPy array of dimension 128×128 for four channels representing FLAIR, T1, T1CE, as well as T2 images. This preprocessing of MR images decreases the computation cost in turn helps to enhance the model prediction accuracy.

Table 1: BraTS Dataset containing Glioma MRI

| Datasets | LGG | Patients (No. of Cases) | HGG |
|---|---|---|---|
| BraTS 2018 | 75 | 285 | 210 |
| BraTS 2019 | 76 | 335 | 259 |
| BraTS 2020 | 76 | 369 | 293 |

B. Data Augmentation

Deep learning models need a lot of data to acquire useful features and to generalize the new data. Data augmentation is a machine learning technique that makes modified copies of existing images to increase the dataset size. It can help deep learning models perform better by increasing the dataset size for the training procedure. It assisted the models to become more accurate and reliable in real-world conditions by enhancing their robustness. Thus, in this work different transformations such as rotation [18], flipping, transpose [24], and cropping [25] are applied and an augmentation ratio of 10% has been implemented on the original images. Figure 1 depicts the process flow diagram for the proposed methods for glioma segmentation.

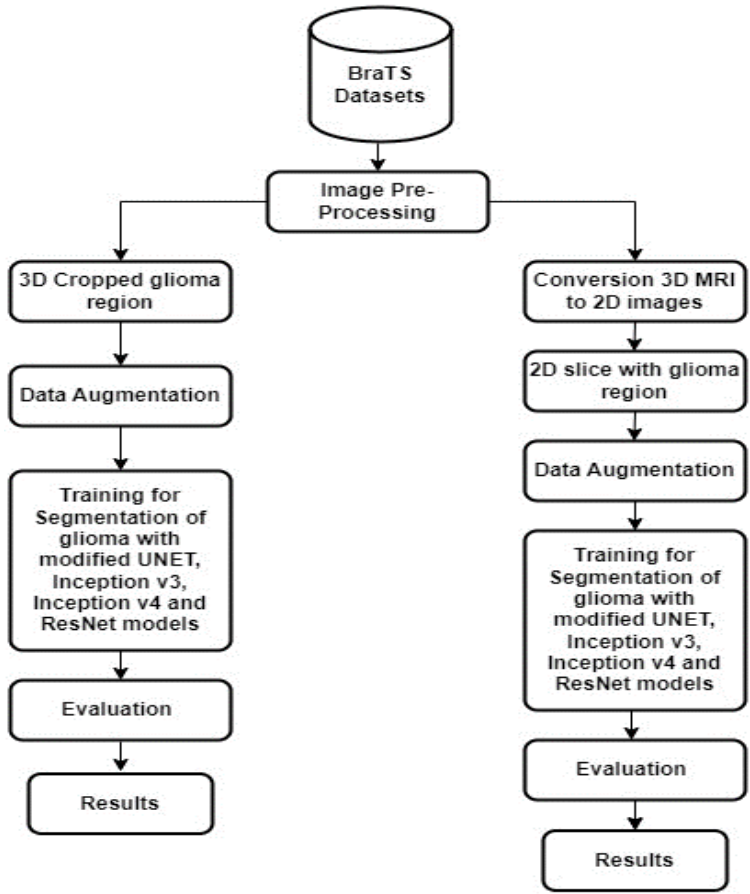

Figure 1: Process flow diagram for the proposed methods for glioma segmentation

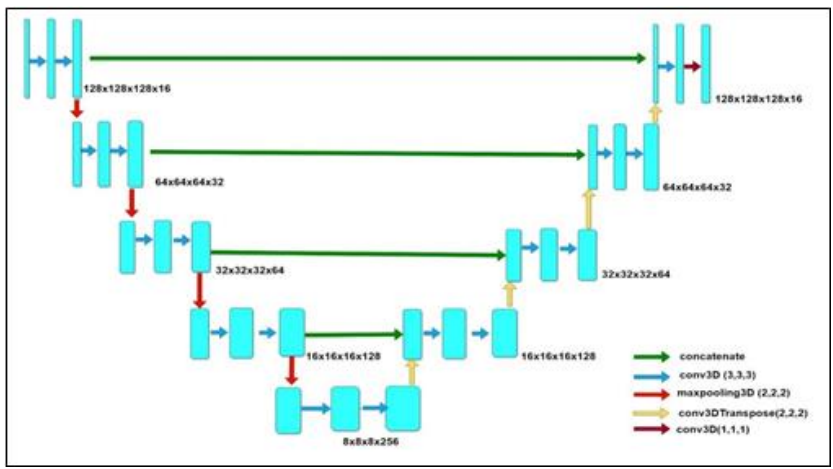

Figure 2: Depicts the modified UNET architecture

C. Deep Learning Network 3D Models

1. Modified UNET Model (UNET): UNET model is a vital Deep Learning (DL) Network 3D Model in the medical area in terms of architecture. It segments the images rooted on pixels originating through convolution layers of the neural network architecture. In this research, a 3D-UNET-rooted model with residual connections [12] is built into it and modification of Vox2Vox's Generator [7] is implemented. The modified UNET model consisted of an encoder and decoder, comprising 15 and 12 convolutional layers, respectively. This proposed model is depicted in Figure 2. The encoder consisted of convolutional layers with filter size 3×3×3 and ReLU (Rectified Linear Unit) activation. The convolutional layers are constructed with increasing feature numbers and the same padding to maintain the same shape collecting essential features from the image. The max pooling layer having a pooling dimension of 2×2×2 followed the convolution layer. This max pooling layer decreased the image size, and finally, the encoder had an image of dimension 8×8×8 with 256 features. The decoder helped bring back the output to size 128×128×128 with four features. The decoder consists of a convolutional layer which is followed through a ConvTranspose Layer having stride as well as pooling size 2×2×2. The decoder decreased the number of features from the convolutional layer and helped build the final output with four extracted features. In the decoder, while decreasing the number of features, some of the accuracies may be lost, so the encoder output from the same level and the decoder output were concatenated to get good accuracy in the output. The dice coefficient of the model increases mainly because of the normalized weight calculated from the segmentation in the BraTS dataset and hybrid loss used: DiceLoss + CategoricalFocalLoss. Hybrid loss is necessary as DiceLoss addresses the imbalance in the background and foreground, and CategoricalFocalLoss deals with the other data imbalance.

2. Modified Pre-trained Inception v3 Model: This Inception v3 model consisted of an Inception Block that was constructed to make an encoder and decoder. This modified architecture has been depicted in Figure 3. Figure 3(a) depicts the Inception Block, and Figure 3(b) depicts the Hybrid Pooling. Figure 3(c) depicts the entire model of the Inception-v3 Model. The Inception Block is designed for feature extraction from the image with the help of different size filters, 1×1×1 and 3×3×3. The output from the convolution layers is concatenated to get the most high-value features from the image. These blocks are connected with hybrid pooling to decrease the dimension and increase the number of features extracted from the image. The hybrid pooling consists of max and average pooling that enhances the model's prediction. The decoder has a convTranspose layer that increases the dimension and decreases the number of features of the final output image. To have better accuracy, the outcome of each inception is concatenated along with the outcome of the inception block in the decoder.

3. Modified Pre-trained Inception v4 Model: The Inception v4 model consisted of an encoder and decoder, which included Inception Block 1 and Block 2. The complete model is shown in Figure 4. Block 1 consists of 7 convolution layers and an average layer. These layers are placed in a way where filter sizes of 1×1×1 and 3×3×3 extract features, and the final was concatenated to give the output image. The Inception Block 2 consisted of convolution layers of filter size 1×1×1, 1×1×7, and 7×7×1, which extracted features from the image and were then concatenated to give the output. All the inception blocks in the encoder were connected with hybrid loss. Hybrid pooling consists of average pooling and max pooling combinations, which are concatenated to decrease image dimension and now pass the relevant features to the next block. In the decoder, the inception blocks are connected with convTranspose, which helped increase the image's dimension and decreased the feature's size. The output from every inception block in the decoder was concatenated from the output of the inception block from the encoder which helped enhance the segmented accuracy.

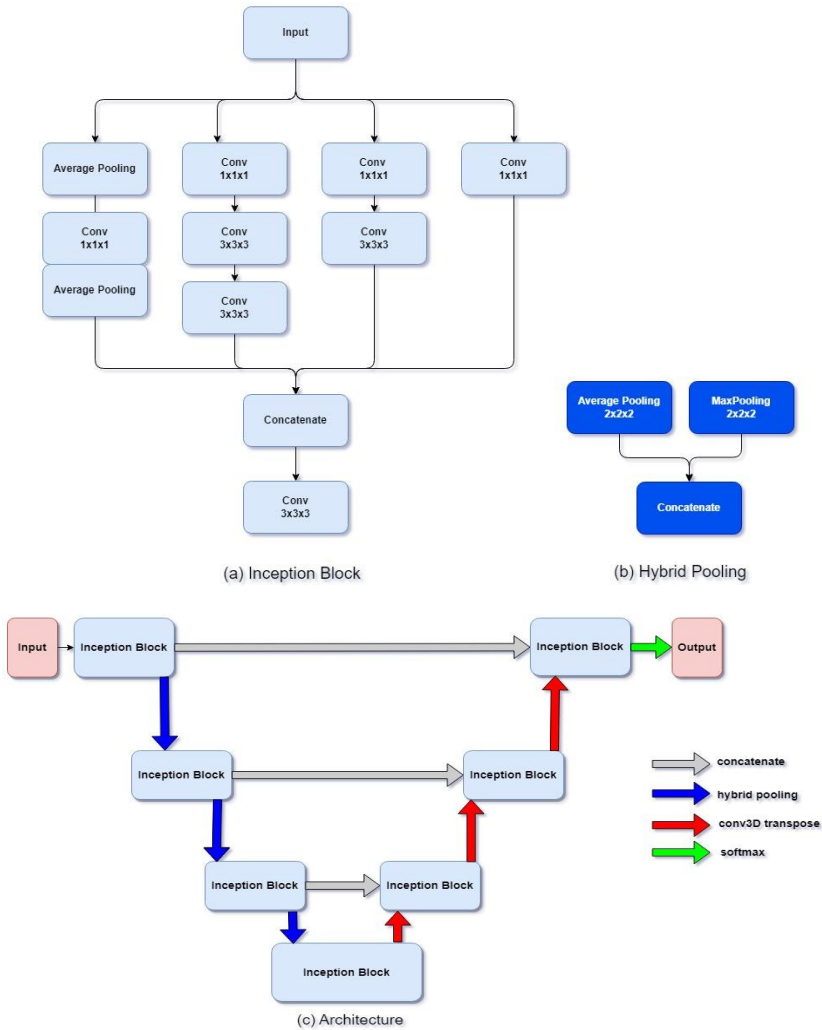

Figure 3: Modified Pre-trained Inception v3 Model (a) Inception Block (b) Hybrid Pooling (c) Complete Model

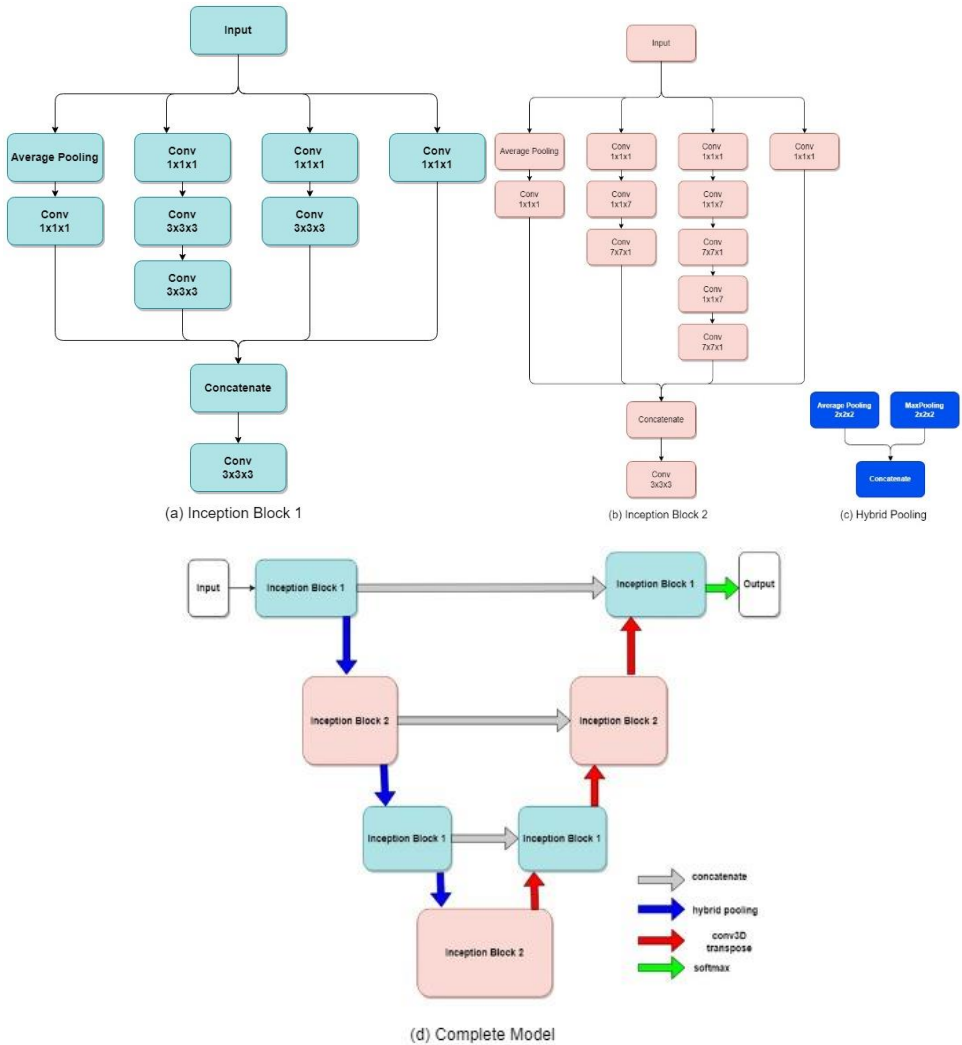

Figure 4: Modified Pre-trained Inception v4 Model (a) Inception Block 1 (b) Inception block 2 (c) Hybrid Pooling (d) Complete Model

4. Modified Pre-trained ResNet Model: The ResNet model has a convolutional layer constructed to form an encoder and decoder to extract the vital features from the image dataset. This dataset was passed through a series of convolution layers of ReLU activations and the same padding to maintain the same dimension. With max pooling, the image changed with the dimension and increased the number of features from the image. The image was then passed to the decoder, where the convTranspose layer increased the image's dimension and decreased the final output's features. In addition, the residual output of every three convolution layers was concatenated with the outcome of the subsequent three convolutional layers in both the decoder and encoder. This helped to extract the necessary features from the imaging dataset to obtain better results.

D. 3D-Model Implementation

The models based on Inception-v3, Inception-v4, and ResNet are implemented in NVIDIA's A100 Tensor Core GPU on the Google Colab Pro+. The model based on UNET was implemented in NVIDIA Tesla K80 GPU with 12GB @ 1.59GHz processors. For the model to work efficiently, the central problem of the dataset had to be solved, which was the imbalance problem where all the features in the segmentation had different weights. To overcome this, normalized weights and DiceLoss with CategoricalFocalLoss were implemented. This solved and addressed the imbalance issue and helped with higher accuracy while training the model.

Table 2: Parameter values implemented on 3D and 2D MRI segmentation of glioma

| Parameter | 3D Value | 2D Value |
| --- | --- | --- |
| Batch Size | 32 | 100 |
| Kernel Initializer | he_uniform | he_uniform |
| Activation Function | ReLU | ReLU |
| Learning rate | 0.0001 | 0.0001 |
| Dropout Layer | 0.2 | 0.8 |
| Final Layer (Activation Function) | Softmax | Softmax |
| Optimizer | Adam | Adam |
| Padding | Same | Same |
| IOU Score Threshold | 0.5 | 0.5 |

The CNNs were arranged into two main categories including the encoder as well as decoder architecture. The model encoder downsizes an image by decreasing the dimension of the image and increasing the vital extracted features. The encoder consists of 3D convolution layers that have a ReLU activation function to prevent exponential computation required to operate the model faster and (3×3×3) filter size for feature extraction. The image was passed through three convolution layers for feature extraction and to other convolution layers with higher feature extraction from the imaging dataset. The dimension was reduced by utilizing the max pooling layer. This step increased the accuracy of extracting the suitable glioma cells in the MRI scan. The resulting image after the encoder was 8×8×8 in dimension with 256 features. The segmentation consists of 4 features with 128×128×128 dimensions. To achieve this, a decoder was used to increase the imaging dataset dimension and decrease the features. The decoder consisted of the 3D convolution networks with decreased features and increased dimension with the ConvTranspose layer. This layer was also fed with the output from the encoder which has identical features and dimensions. This increased the learning precision by collecting the data samples in the particular glioma area. The parameter table for 3D models is shown in Table 2.

E. 2D-Model Implementation

The BraTS dataset is 240×240×155 in dimension, representing the image's height, width, and depth. For the 2D model, the input would only be a 2D NumPy array, so the database had to go through preprocessing, which included normalizing the dataset and cropping the image, only to take the essential part in focus and convert the 3D image into a 2D image.

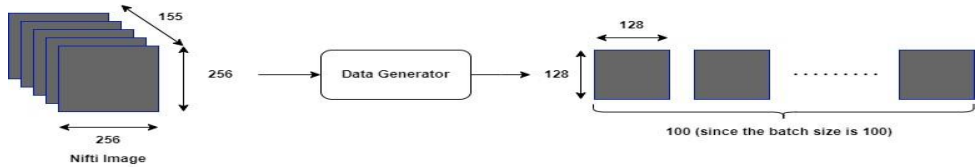

Figure 5: Conversion from 3D model to 2D showcasing the Nifti image sizes and output of the Data Generator along with the batch size

1. Conversion of 3D model to 2D model: The BraTS datasets consist of the Nifti images of the shape's T1, T1CE, T2, and FLAIR layers (240,240,155), which was too large and would increase the model computation cost. The segmented mask available in the dataset is non-uniform, which was made uniform with the help of the MinMaxScalar function in Scikit-learn [26]. The images were cropped to (128,128,128). This shape helped for faster and easier training. Once the non-uniformity and large-size problems were resolved, the image was still in 3D. To convert the 3D to 2D images, from each Nifti image, every slide in it was considered. Each slide is 2D in shape which is our requirement. The block diagram in Figure 5 shows the conversion from 3D to 2D.

While converting a 3D model to a 2D model, the primary task was to make a data generator that gives a 2D NumPy array of the desired batch size. This is achieved by cropping the image to the shape of (128×128×128) and stacking all the T1, T1CE, T2, and FLAIR layers into one. This results in creating (128×128×128×4) shaped NumPy arrays as input to the architecture. This proposed architecture is made into 2D through the replacement of all the 3D layers used in the 3D model with a 2D model. The architecture was kept the same, consisting of an encoder and decoder. The loss function that was used in the proposed architecture is a combination of categorical focal loss as well as dice loss. The 3D model gave better results with much more efficiency because all the T1, T1CE, T2, as well as FLAIR images, were stacked together with the help of the NumPy. That facilitates the framework to determine the various patterns in the imaging dataset better and thus results in a very efficient training of the model resulting in an accurate predicting segmentation model.

The data generator is the critical feature that converts the image from 3D to 2D. This was achieved by taking an empty NumPy array of dimension layer 128×128 with four channels for both the image and segmentation. A Nifti image consists of a 240×240 dimension and 155 slides together above each other in the BraTS dataset. Therefore, every 240×240-dimension slide is assigned to an empty NumPy array and used in batches for model training. The 2D convolutional neural networks are arranged into two main parts, the encoder and the decoder. The encoder downsizes an image by decreasing the image size and increasing the extracted features. The encoder consists of 2D convolution layers with a ReLU activation function to prevent exponential computation required to operate the model faster and (3×3) filter size to extract the features. The image dataset is translated via three convolutional layers for extraction of vital features and to other convolutional layers with higher features mined from the imaging data. The dimension was reduced utilizing the max pooling layer. This step increased the accuracy of extracting the correct glioma cells in the MRI scan. The resulting image after the encoder is 8×8 in dimension with 256 features. The segmentation in the BraTS dataset consists of 4 features with 128×128 dimensions. To achieve this, a decoder is used to increase the image dimension and decrease the number of features. The decoder consists of the 2D convolution networks with decreased features and increased dimension with the convTranspose layer. The conv2Dtranspose layer is also fed with the output from the encoder model having identical dimensions as well as features. This increased the learning precision by collecting the data samples in the particular brain glioma area. The parameter table for 2D models is shown in Table 2.

The 2D models used for BraTS image segmentation are made with the help of the Keras layers, which consist of the following layers. The Conv2D layer acts as a 2-dimensional convolutional layer to learn the local patterns in the input NumPy array. The MaxPooling2D is a 2D layer that was utilized for downsampling the input NumPy array. This was done by applying a pooling window to the input NumPy array and selecting the maximum value within the window as the output value. The Upsampling2D is a 2D deconvolutional layer that was used to up-sample the input NumPy array. This was carried out by applying filters to the input NumPy array and using the resulting feature maps to up-sample the input NumPy array. The dropout layer arbitrarily sets a defined percentage of the applied input units to zero during training, which reduces the complexity of the model and forces it to rely more on the remaining units. This, in turn, helped the model to become more efficient in predicting unseen data and reduced the model computation. The Concatenate layer concatenates multiple input tensors along a specific axis.

Four 2D and 3D models were implemented in the above layers to get high dice-scoring models. The models are UNET, Inception-v3, Inception-v4, and ResNet. These models had a consistent architecture having an encoder as well as a decoder. This encoder model was responsible for encoding the input dataset into a compressed representation, and the decoder model was accountable for decoding the compact representation back into the original form. The advantage of using an encoder-decoder architecture was that it allowed the proposed model to predict a compact representation of the applied input dataset that captured the most important features and patterns. This compact representation helped the model to be more efficient and easier to train.

F.  K-fold Cross-Validation for Model Evaluation

There is selected the K-Fold Cross Validation method to assess the performance and generalizability of the models. When there is a lack of data, the dataset is unbalanced, or sampling bias is a concern, this technique helps in statistical modeling and machine learning tasks. The dataset was split into five train-test folds or splits at random, and the average of these assessments was utilized to calculate the proposed model performance measures. The use of this technique reduces the possibility of overfitting to a particular train-test split by choosing the model that produces consistent results across many subsets.

G. Evaluation Measures

Based on the following criteria, the proposed model's performance has been evaluated utilizing the Dice coefficient, Hausdorff distance, and Accuracy. To determine the overlap amongst the predictions as well as the ground truth, the Dice score is employed. The Hausdorff score determines the effective distance between the predicted area (Y) boundaries and that of the actual ground-truth measure (X) masks. Equations (1) and (2) show the evaluation metrics:

$$Dice\ Score = \frac{2 \times TP}{(TP+FP)+(TP+FN)} \quad (1)$$

$$Hausdorff\ Distance = \left(\frac{1}{X}\sum_{x \in X} \min_{y \in Y} d(x,y) + \frac{1}{Y}\sum_{y \in Y} \min_{x \in X} d(x,y)\right)/2 \quad (2)$$

Wherein TP is the abbreviation for True Positive, TN represents True Negative values, further, FP illustrates the False Positive values, and FN represents the False Negative to the Dice Score and (x,y) stands for the points on the margins of the predicted image to the Hausdorff Distance as depicted in Figure 6.

H. Loss Function for Brain Glioma Segmentation Task

In this work, the Dice Score often known as F1 score loss was employed to quantify the likeness of the forecasted mask as well as the ground truth mask of the glioma on a per-pixel basis. The Dice Score values 0 and 1 denote no overlap and a perfect match, respectively and it addresses class imbalance issues by giving the foreground and background classes equal weight (Milletari, Navab and Ahmadi, 2016). Dice Score is calculated using the intersection of the two masks and their total area. The formula is given as:

$$Dice\ Score = \frac{2 \times TP}{(TP+FP)+(TP+FN)} \quad (3)$$

$$Dice\ Loss = 1 - Dice\ Score \quad (4)$$

Further, the Dice Loss is combined with Cross-Entropy Loss (CE) or Focal Loss (FL), to form a multi-objective loss. This combination allows the model to capture both localization accuracy (e.g., cross-entropy) and boundary refinement (e.g., Dice loss) simultaneously. CE tackles the class imbalance issues by assigning large weights to the easy and hard misclassified instances (for example, background having a noisy texture or another partial object, etc.) and to downside-weight easy instances (for example, the Background objects) as depicted in Figure 7.

$$Focal\ Loss = -\sum_{i=1}^{i=n}(1-p_i)^\gamma \log b(p_i) \quad (5)$$

where $p$ is the suggested framework evaluated probability and $\gamma$ is the tunable focusing parameter.

In the proposed models, a grouping of Dice Loss as well as Categorical Focal Loss has been implemented. By incorporating the Dice loss into a comprehensive loss function.

$$Total\ Loss = Dice\ Loss + (1 * Focal\ Loss) \quad (6)$$

The implementation used equal weight (weight = 1) which means giving equal importance to both the dice loss and the focal loss. This is a balanced approach where both loss terms contribute equally to the overall loss function. Thus, the model can effectively balance different aspects of brain glioma segmentation.

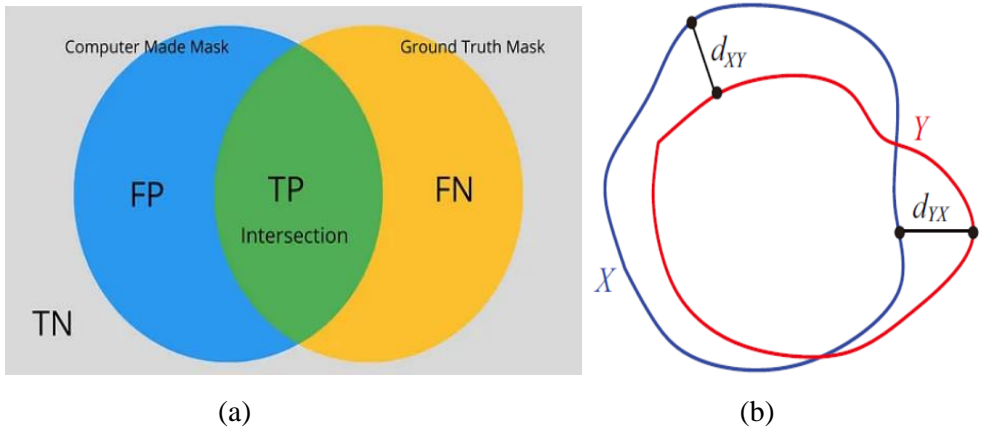

(a)                (b)

Figure 6: (a) Dice Score – The intersection of forecasted and ground truth masks (b) Pictorial representation of Hausdorff Distance

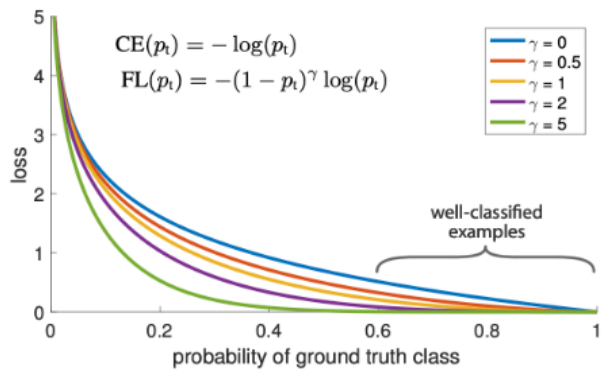

Figure 7: Focal Loss adds a factor $(1 - p_t)^\gamma$ to the standardized cross-entropy measure. Setting γ>0 minimizes the comparative losses for well-categorized instances (p_i " " >" 0.5"), giving a high focus on hard, and misclassified instances.

## 4. Results and Discussion

A. 2D and 3D Models Segmentation Visualizations

Visualization of segmented glioma images is pivotal in clinical practice, research, and patient education. These images provide clinicians with crucial diagnostic information, aiding in accurate tumor delineation and treatment planning by identifying target areas while minimizing damage to healthy tissues. Furthermore, segmented glioma images facilitate disease monitoring over time, allowing for adjustments to treatment plans based on changes in tumor characteristics and response to therapy. Beyond clinical applications, these images empower patients by enhancing their understanding of their condition and treatment options, fostering active involvement in their healthcare decisions. Additionally, segmented glioma images serve as valuable resources for research and education, supporting the development of new diagnostic and treatment techniques and providing real-life examples for training healthcare professionals in interpreting medical imaging data accurately. Figure 8 depicts the 2D Segmentation outputs for BraTS (a) 2018; (b) 2019; (c) 2020 datasets; Row 1: Original MRI sequences images datasets (FLAIR, T1, T1-CE, T2); Row 2: UNET, Row 3: Inception v3, Row 4: Inception v4, Row 5: ResNet. Figure 9 illustrates the 3D Segmentation outputs for BraTS (a) 2018; (b) 2019; (c) 2020 datasets; n=55; Row 1: Original MRI sequences images datasets (FLAIR, T1, T1-CE, T2); Row 2: UNET, Row 3: Inception v3, Row 4: Inception v4, Row 5: ResNet.

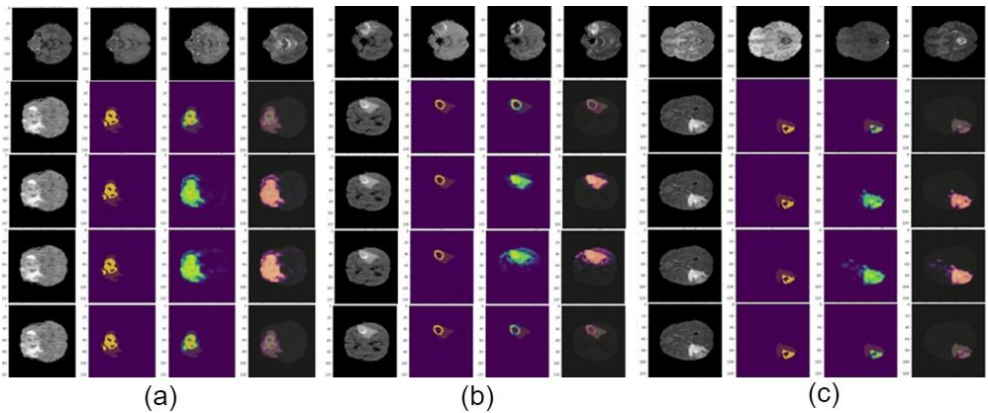

Figure 8: 2D Segmentation outputs for BraTS (a) 2018; (b) 2019; (c) 2020 datasets; Row 1: Original MRI sequences images datasets (FLAIR, T1, T1-CE, T2); Row 2: UNET, Row 3: Inception v3, Row 4: Inception v4, Row 5: ResNet.

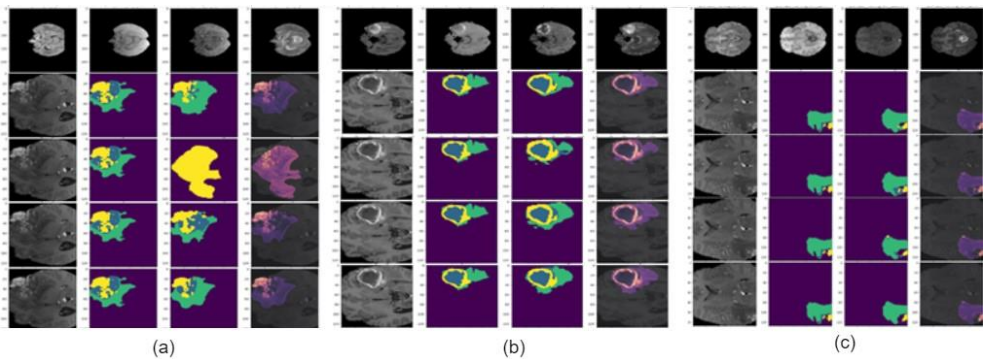

Figure 9: 3D Segmentation outputs for BraTS (a) 2018; (b) 2019; (c) 2020 datasets; n=55; Row 1: Original MRI sequences images datasets (FLAIR, T1, T1-CE, T2); Row 2: UNET, Row 3: Inception v3, Row 4: Inception v4, Row 5: ResNet

B. 2D and 3D Models Segmentation Outcomes

Evaluation of the proposed models over the BraTS datasets reveals superior accuracy as well as performance compared to previous challenge-winning methods, attributed to hyperparameter fine-tuning and data augmentation techniques applied to UNET and pre-trained models like Inception v3, v4, and ResNet. These models, pre-trained on ImageNet, leverage their broad visual feature understanding, and further fine-tuned on glioma MRI data to efficiently capture intricate patterns, enhancing feature representation and segmentation precision.

Table 3 and Table 4 show the evaluation results of different models for BraTS 2018, BraTS 2019, as well as BraTS 2020 datasets for 2D as well as 3D MRI. Table 5 gives the comparative analysis between the proposed methods for 2D and 3D segmentation and other baseline approaches that won the BraTS Challenges on BraTS 2018, BraTS 2019, as well as BraTS 2020 datasets, respectively. It is observed that the proposed technique has achieved accuracy scores higher than the winners of the BraTS Challenges. This is attributed to the hyperparameter fine-tuning and the data augmentation techniques applied to UNET and each of the pre-trained models. It is built to get both local and global insights from the 2D and 3D MR images. The skip connections help in preserving the spatial data which is crucial for precise segmentation of the different regions of the glioma by combining features from different scales. Inception v3, Inception v4, and ResNet have been pre-trained on the ImageNet dataset and were particularly well-suited due to their broad understanding of visual features. Further, fine-tuning on MRI of glioma data for segmentation leveraged this pre-trained knowledge. Also, these modules helped in efficiently utilizing computational resources and performing convolutions at multiple scales concurrently. Better feature representation was seen in the enhanced architectures, which was beneficial for capturing the diverse and intricate patterns present in glioma images.

ResNet's deep residual architecture with skip connections enabled the training of very deep networks. This is particularly useful for learning hierarchical features in glioma images, where the complexity of patterns in the glioma structure needs such a model. The residual connections mitigate the vanishing gradient problem, facilitating stable and efficient training even in deep networks. As seen in Table 5, the modified architectures have been compared with the winners of the BraTS Challenges 2018, 2019, and 2020. Various parameters such as the Dice Score, Loss Function, Optimizer, and number of epochs have been compared. L(dice) denotes the soft dice loss, L(L2) represents loss on the variational auto-encoder (VAE) branch output, L(KL) is the standard VAE penalty term, L(CF) is the Categorical Focal Loss, B18 denotes the BraTS 2018 dataset, B19 denotes the BraTS 2019 dataset, and B20 is the BraTS 2020 dataset. The time taken for one epoch is complete is also shown in Table 5.

Table 3: Evaluation metrics of models implemented on 2D MRI.

| Performance Metrics | Accuracy | | | Hausdorff | | |
|---|---|---|---|---|---|---|
| Datasets | 2018 | 2019 | 2020 | 2018 | 2019 | 2020 |
| | 0.9974 | 0.9731 | 0.9769 | 3 | | |
| | 0.9234 | 0.9289 | 0.9902 | 3 | | |
| | 0.9974 | 0.9734 | 0.9535 | 3 | | |
| | 0.9977 | 0.9734 | 0.9831 | 3 | | |

| 2D Models | Dice Score – Performance Metrics Performance | | | | | |
|---|---|---|---|---|---|---|
| | Datasets | | | | | |
| | 2018 | 2019 | 2020 | 2018 | 2019 | |
| UNET | 0.7943 | 0.7981 | 0.8234 | 3 | 3 | |
| Inception-v3 | 0.7860 | 0.7859 | 0.7855 | 3 | 3 | |
| Inception-v4 | 0.7933 | 0.7867 | 0.7865 | 3 | 3 | |
| ResNet | 0.8112 | 0.8299 | 0.8312 | 3 | 3 | |

Table 4: Evaluation metrics of models implemented on 3D MRI.

| Hausdorff Performance Metrics | | | Accuracy - Performance Metrics | | |
|---|---|---|---|---|---|
| Datasets | | | Datasets | | |
| 2019 | 2020 | | 2018 | 2019 | 2020 |
| 10.625 | 13.0715 | | 0.9880 | 0.9887 | 0.9871 |
| 11.1141 | 14.3513 | | 0.9786 | 0.9817 | 0.9747 |
| 11.1141 | 14.3513 | | 0.9785 | 0.9817 | 0.9747 |
| 10.8453 | 11.7000 | | 0.9891 | 0.9869 | 0.9878 |

| 3D Models | Dice Score – Performance Metrics–Performance | | | Datasets | | |
|---|---|---|---|---|---|---|
| | | 2018 | 2019 | 2020 | 2018 | |
| UNET | | 0.9877 | 0.9884 | 0.9869 | 12.6112 | |
| Inception-v3 | | 0.9784 | 0.9807 | 0.9733 | 14.8291 | |
| Inception-v4 | | 0.9782 | 0.9807 | 0.9733 | 14.8600 | |
| ResNet | | 0.9888 | 0.9866 | 0.9875 | 10.8977 | |

Table 5: Comparative analysis of the proposed models for 2D and 3D methods which won the BraTS Challenges of 2018, 2019, and 2020

| | Brats 2018 Winner | Brats 2019 Winner | Brats 2020 Winner | UNET | Inception-v3 | Inception-v4 | ResNet |
|---|---|---|---|---|---|---|---|
| Image size | 160× 192× 128 | 4× 128× 128× 128 | (128× 128× 128) | 128× 128× 128× 4 | 128× 128× 128× 4 | 128× 128× 128× 4 | 128× 128× 128× 4 |
| Encoder endpoint | 256× 20× 24× 16 | 128× 16× 16× 16 | - | 8×8×8×256 | 8×8×8×256 | 8×8×8×256 | 8×8×8×256 |
| Dice Score | 0.8839 | 0.88796 | 0.91148 | B18-0.9877 B19-0.9884 B20-0.9869 | B18-0.9784 B19-0.9807 B20-0.9733 | B18-0.9782 B19-0.9807 B20-0.9733 | B18-0.9888 B19-0.9866 B20-0.9875 |
| Loss Function | L(dice) + 0.1 ∗ L(L2) + 0.1 ∗ L(KL) | L(dice) | L(dice) | L(dice)+ L(CF) | L(dice)+ L(CF) | L(dice)+ L(CF) | L(dice)+ L(CF) |
| Optimizer | Adam | Adam | Ranger | Adam | Adam | Adam | Adam |
| Learning rate | 1e−4 | 1e−4 | 1e−4 | 0.0001 | 0.0001 | 0.0001 | 0.0001 |
| Hausdorff Distance(mm) | 4.4834 | 4.61809 | 4.3 | B18-12.6112 B19-10.625 B20-13.0715 | B18-14.8291 B19-11.1141 B20-14.3513 | B18-14.86 B19-11.1141 B20-14.3513 | B18-10.8977 B19-10.8453 B20-11.7 |
| No of Epochs | 300 | 405 | 400 | 100 | 100 | 100 | 100 |
| Time Taken | 9 min | - | - | 4min | 4min | 4min | 4min |

| for one epoch | | | | | | | |
|---|---|---|---|---|---|---|---|

C. 2D and 3D Models Comparative Graphical Results

The graphs in Figure 10 and Figure 11 show the comparative curves for accuracy and loss for the modified architecture of UNET and pre-trained models Inception-v3, Inception-v4, and ResNet for the 3D models on BraTS datasets 2018, 2019, and 2020.

As seen in Figures 10 and 11, the ResNet model performs better than the other models closely followed by the UNET. This model is well-suited for glioma segmentation due to its ability to capture both local and global features while utilizing skip connections for spatial context preservation and leverage pre-training for improved generalization. When comparing the different models, the choice among these models may depend on factors namely the availability of data, computational assets, and specific characteristics of the glioma imaging dataset. Fine-tuning these models on glioma imaging data is often necessary to adapt them to the nuances of glioma segmentation.

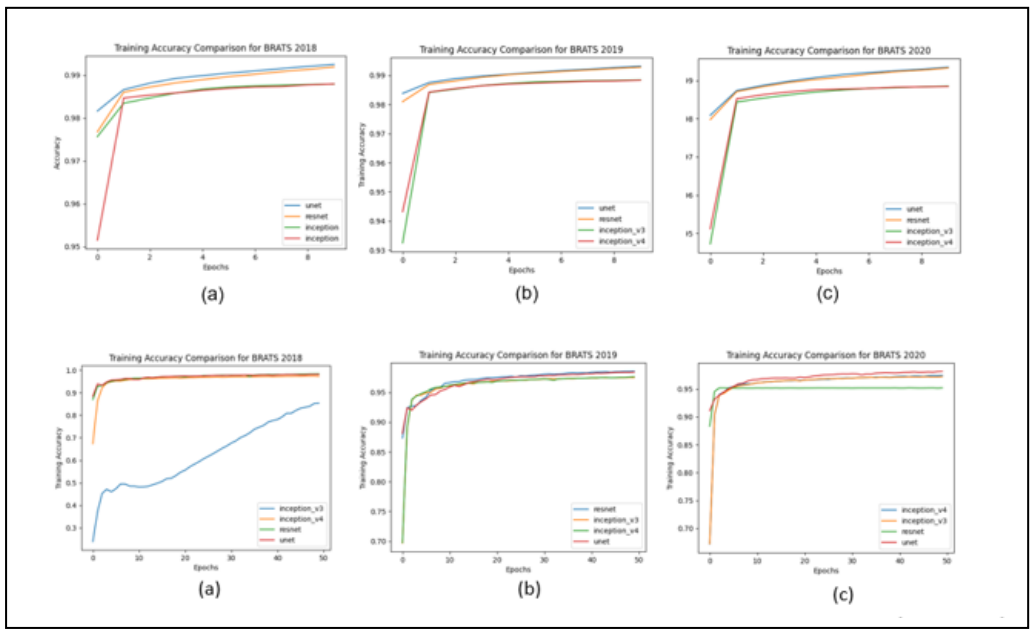

Figure 10: Accuracy curves for 2D models - top row as well as 3D models – bottom row. (a) Brats 2018 (b) BraTS 2019 (c) BraTS 2020.

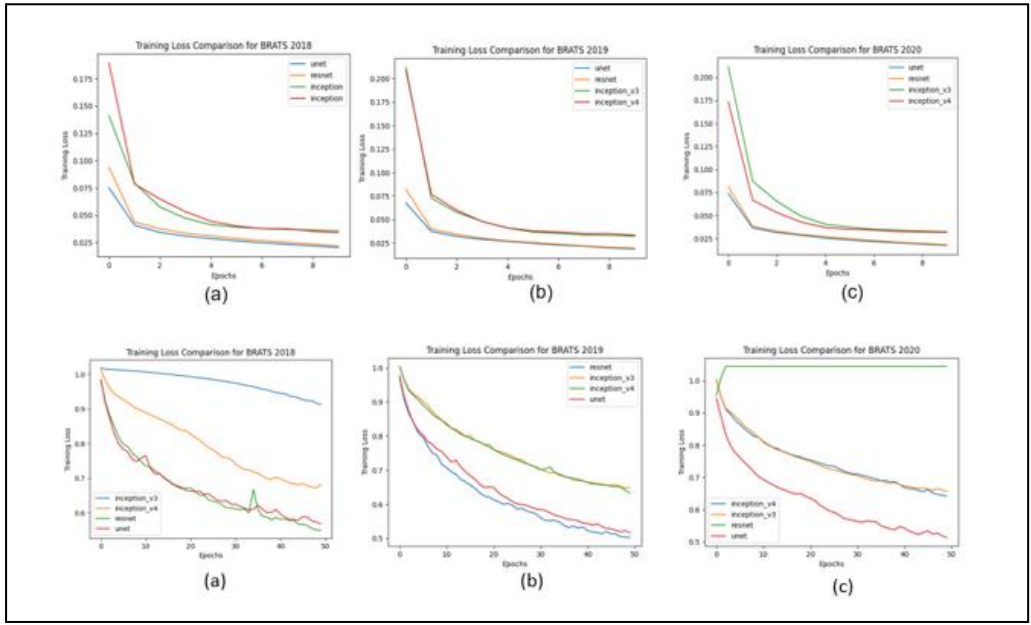

Figure 11: Loss curves for 2D models - top row as well as 3D models – bottom row. (a) Brats 2018 (b) BraTS 2019 (c) BraTS 2020.

## 5. Conclusion

In conclusion, developing a completely automated method for glioma segmentation is required for precise treatment planning and prognosis. For the segmentation of gliomas, contemporary methods that employ fully convolutional neural networks (FCNs) have produced encouraging results. Due to their high memory requirements and high computational costs, choosing between 2D and 3D convolutions is still difficult. This work used models for glioma segmentation based on UNET, Inception, and ResNet, which, according to thorough research on the BraTS 2018, BraTS 2019, as well as BraTS 2020 datasets, demonstrated higher potential for effective segmentation. The ResNet model has demonstrated impressive performance in segmenting brain images, achieving an accuracy of 98.91% for 3D segmentations and an even higher accuracy of 99.77% for 2D segmentations. This indicates the proposed framework's remarkable capability to correctly determine and delineate structures of interest within both 3D volumetric data and 2D image slices. Moreover, the dice scores, that determine the overlap between the model's segmentation as well as the ground truth level, further underscore its effectiveness. For 2D segmentations, the model achieved a dice score of 0.8312, demonstrating a substantial agreement between the model's predictions as well as the actual segmentation. In the case of 3D segmentations, the dice score notably increased to 0.9888, signifying an even higher level of concordance between the model's output and the true segmentation. These results highlight the robustness and accuracy of the ResNet model in performing both 2D and 3D segmentations of brain images, making it a valuable tool for various medical imaging applications, including tumor detection, treatment planning, and disease monitoring. These findings suggest that the suggested methodologies could greatly increase the precision and efficacy of glioma segmentation and enable more efficient treatment planning for brain glioma patients.